\documentclass[aps,prb,
twocolumn, 
               a4paper,amsmath,amssymb,superscriptaddress]{revtex4-1}

\usepackage[utf8]{inputenc}
\usepackage[english]{babel}
\usepackage{graphicx}
\usepackage{feynmp}
\usepackage{bm}
\usepackage{color}
\usepackage{braket}

\newcommand{\e}{ {\rm e}}
\newcommand{\ET}{ $\alpha$-(BEDT-TTF)$_2$I$_3$}

\newcommand{\bk}{ \bm{k}}
\newcommand{\bkD}{ \bm{k}_{\rm D}}

\newcommand{\tH}{\tilde{t}}
\newcommand{\tomega}{\tilde{\omega}}
\newcommand{\x}{\tilde{x}}
\newcommand{\y}{\tilde{y}}
\newcommand{\z}{\tilde{z}}
\newcommand{\HH}{{\rm H}}
\newcommand{\LL}{{\rm L}}

\newcommand{\bZ}{\bar{Z}}

\newcommand{\bq}{\bm{q}}
\newcommand{\g}{\gamma}

\newcommand{\sn}{{\sigma _{\nu}}}

\begin{document}
\title{
Electric Transport of Nodal Line Semimetal in 
Single-Component Molecular Conductors 
}

\author{Yoshikazu \surname{Suzumura}}
\email{Correspondence:suzumura@s.phys.nagoya-u.ac.jp(Y.S.)} 
 \affiliation{Department of Physics, Nagoya University,
             Nagoya 464-8602, Japan}            

\author{Reizo \surname{Kato}}
\email{reizo@riken.jp} 
 \affiliation{RIKEN, 2-1 Hirosawa, Wako-shi, Saitama 351-0198, Japan}

\author{Masao \surname{Ogata}}
\email{ogata@phys.s.u-tokyo.ac.jp} 
 \affiliation{Department of Physics, University of Tokyo, Bunkyo, Tokyo 113-0033, Japan}

\begin{abstract}
We examine an effect of acoustic phonon scattering on an electric conductivity 
 of  single-component molecular conductor  [Pd(dddt)$_2$] 
(dddt = 5,6-dihydro-1,4-dithiin-2,3-dithiolate) with a half-filled band 
 by applying  the previous calculation 
in a two-dimensional model with   Dirac cone 
[Phys. Rev. B {\bf 98},161205 (2018)], where
 the electric transport by the impurity scattering 
  exhibits the noticeable
  interplay of the Dirac cone 
 and the phonon scattering, 
resulting in  a maximum of the conductivity with increasing temperature.
The conductor shows a  nodal line semimetal, 
 where the band crossing of HOMO (Highest Occupied Molecular Orbital) and 
LUMO (Lowest Unoccupied Molecular Orbital)
 provides a loop of Dirac points  located close to the Fermi energy 
 followed by the density of states (DOS) similar to that of 
 two-dimensional Dirac cone. 
Using a tight-binding (TB) model [arXiv:2008.09277], 
which was  obtained 
 usig the crystal structure observed 
 from a recent  X ray diffraction experiment under pressure, 
 it is shown that the obtained conductivity  explains reasonably 
 the anomalous behavior in 
  [Pd(dddt)$_2$] exhibiting  
 almost  temperature independent resistivity at finite temperatures.
 This paper demonstrates  a crucial  role of 
  the acoustic phonon scattering at finite temperatures in the electric conductivity  
   of  Dirac electrons. 
 The present theoretical results of conductivity 
are compared with those of experiments.  
\end{abstract}

\maketitle

\section{Introduction} 

In molecular solids,  various electronic  properties 
 from metallic to insulating states have been examined due to the interplay of 
transfer energies and mutual interactions between molecules.\cite{Seo2004} 
However there is another electronic state, where the temperature dependence 
 of the resistivity 
 does not show  either metallic nor insulating behavior.
It originates  from a band structure
 of a massless Dirac electron,~\cite{Herring1937,Fu2007}
  where the conduction  and valence bands cross at a certain momentum 
 in the Brillouin zone.
 Such a state  has been  discovered 
  in  the two-dimensional materials of  graphene\cite{Novoselov2005_Nature438} and  organic conductors.
\cite{Katayama2006_JPSJ75,Kajita_JPSJ2014} 
This  state has been extensively studied in three-dimensional system to show 
 a nodal line semimetal  i.e., a loop   of Dirac points 
 in inorganic conductors~
\cite{Murakami2007,Hirayama2018,Bernevig2018}
 and in molecular conductor.~\cite{Kato_JACS} 

 The Dirac electron in organic conductor with a zero-gap state was found in 
   two-dimensional   \ET 
      (BEDT-TTF=bis(ethylenedithio)tetrathiafulvalene), 
 \cite{Katayama2006_JPSJ75,Kajita_JPSJ2014} 
       using  a tight-binding (TB) model 
     with transfer energies estimated by 
         the extended H\"uckel method.\cite{Kondo2005} 
It should be noted that  the TB model~\cite{Mori1884,Kino1996}
  describes successfully  the electronic states of 
the molecular conductors.
These Dirac electrons 
  are studied experimentally and theoretically to comprehend  
   physical properties  in the  bulk  system.
Noticeable temperature dependence of 
 Hall coefficient,\cite{Kobayashi2008,Tajima2012}
  NMR~\cite{Katayama2009_EPJ,Takahashi2010,Hirata2016},
 anisotropic conductivity,~\cite{Suzumura_Igor_2014} 
 and Nernst coefficient~\cite{Igor_Ogata_2013}
 have been  obtained 
 since the Dirac point is located close to the chemical potential.

Another  molecular Dirac electron system  was found in  a single-component molecular conductor,  [Pd(dddt)$_2$] 
(dddt = 5,6-dihydro-1,4-dithiin-2,3-dithiolate).\cite{Kato_JACS} 
 The application of pressure is useful for modification of conducting properties in single-component molecular crystals with soft lattices. 
\cite{2009,2016,2018,2017,2015}
  [Pd(dddt)$_2$] 
 under high pressure  exhibits     
  almost temperature independent  resistivity.\cite{Cui} 
First-principles calculations  indicate that this materials 
belongs to the three-dimensional Dirac electron system,
\cite{Tsumuraya_PSJ_2014}
  consisting of HOMO (Highest Occupied Molecular Orbital) and LUMO (Lowest Unoccupied Molecular Orbital) bands, and a TB model 
 exhibits  a loop of Dirac points called  
      a  nodal line semimetal.\cite{Kato2017_JPSJ} 
There are several studies  on the effective 
Hamiltonian, where a general two-band  model is introduced\cite{Liu2018}
 and the explicit calculation is performed 
 for  the nodal line semimetal.
\cite{Tsumuraya2018_JPSJ}

However  the conductivity  is not yet clearly understood 
 although 
 the almost temperature independent conductivity is believed 
 to be an evidence of Dirac electrons.
  From the theoretical point of view, the conductivity at the zero doping 
  shows a universal conductance at absolute temperature\cite{Novoselov2005_Nature438}
  but increases linearly with increasing temperature due to the linear 
 increase  of the density of states,\cite{Katayama2006_cond,Kobayashi2008}
 when  the temperature is larger  than the energy of the damping  
     by the impurity scattering. 
 Such an increase  is incompatible with the experiment of 
 the almost constant resistivity in two-dimension.\cite{Tajima2007_EPL}
 The problem comes from the fact that the constant behavior can be 
 obtained only for the  damping energy 
 being larger than the temperature, which 
 is not the case of  the molecular conductors.
Recently, to comprehend the almost constant conductivity, 
 a possible mechanism 
   has been proposed for two-dimensional 
      Dirac electrons, where  
the scattering by an acoustic phonon 
  plays the crucial   role.
\cite{Suzumura2018_PRB}

The conductivity of [Pd(dddt)$_2$] 
 showing  almost temperature independent behavior,\cite{Kato_JACS} 
  was examined theoretically\cite{Suzumura2017_JPSJ,Suzumura_Kato2017_JPSJ}
 using a TB model of  the H\"uckel calculation for  a crystal 
 structure obtained by first-principles calculation under pressure.
 The DOS  exhibits a linear dependence close to the chemical potential  but 
 the region for the relevant energy is narrow compared with that for the constant resistivity.
The conductivity in terms of the above TB model 
 shows a large anisotropy and the almost constant behavior
   at high temperature, 
 since the reduction of the DOS for larger energy suppresses  the increase of 
 the conductivity. 
In this case, the TB model suggests  
 the almost constant resistivity in [Pd(dddt)$_2$] 
 due to the decrease of the  DOS, but not by 
  the property relevant to the  Dirac cone.
Recently a TB model was reexamined using the crystal structure, 
 which was obtained under high pressure.\cite{Kato2019_TB}
Surprisingly,  this band calculation shows the DOS, which depends  
linearly  in a wide region of the energy being 
 compatible with temperature  of almost constant conductivity.
Thus, we reexamine  the almost constant resistivity in [Pd(dddt)$_2$] 
 using such a newly found TB model and by taking acount of   
   the acoustic phonon scattering. 
The present paper demonstrates that such a mechanism does exists 
 as the evidence of Dirac electrons for single-component molecular conductor.
Note that three-dimensional system with nodal line of Dirac points is important compared with two-dimensional systems with a Dirac point, since the almost constant conductivity is obtained by the combined effect  of the nodal line and  the acoustic phonon.

In Sect. 2,  based on the  recently obtained  TB model under pressure,
the nodal line and the DOS are calculated to find 
   a wide region for the linear  dependence of the DOS. 
 A formulation  for the conductivity is given by taking account of 
  both the impurity and  electron-phonon (e-p) scatterings. 
In Sect. 3, by  calculating the  chemical potential for the half-filled band,  
 we examine the temperature dependence of the anisotropic conductivity 
 and show the almost constant conductivity for  reasonable choices 
 of the e-p coupling constant. 
The corresponding resistivity is compared with that of the experiment.
In Sect. 4,  summary and discussion are given.

\section{Model and Formulation}

\subsection{TB model}

\begin{figure*}[tb]
\begin{center}
  \includegraphics[width=0.9 \linewidth]{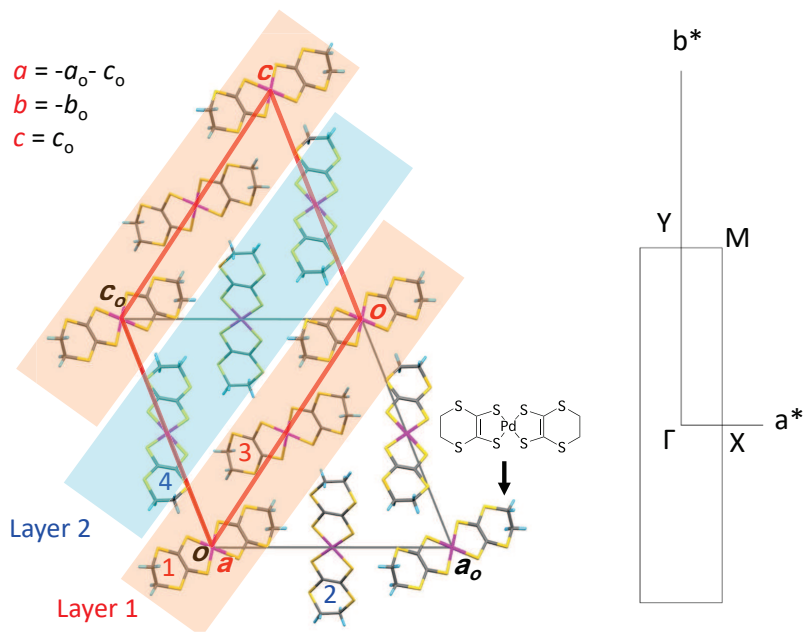}
\end{center}
\caption{
(Color online)
Crystal structure of [Pd(dddt)$_2$] shown in the $ac$ plane, 
where the molecule is stacked along the $b$ direction perpendicular to the plane. 
Layer 1 (molecules 1 and 3) and Layer 2 (molecules 2 and 4) 
 are parallel to the $ab$ plane 
and alternated along the $c$ direction.
Lattice vectors  $\bm{a}$, $\bm{b}$,and  $\bm{c}$ form  
 a  new unit cell, where  $\bm{a}$ is introduced 
 for the simple representation of
the energy band (see the main body).
 The location of the Pd atom in the unit cell   
 is given by ($x$,$y$,$z$)=(0,0,0), (1/2,1/2,1/2),(1/2,1/2,0)
and (1/2,0,1/2) for molecules 1, 2, 3, and 4, respectively. 
The right figure denotes the Fourier space, where $\bm{a}^*$ and $\bm{b}^*$ are 
 reciprocal lattice vectors and $\Gamma$, X, Y, and M are TRIM (time reversal invariant 
 momentum).
} 
\setlength\abovecaptionskip{0pt}
\label{fig1}
\end{figure*}

\begin{table}
\caption{ 
Transfer energies for $P$=5.9GPa,\cite{Kato2019_TB}
 which are multiplied by $10^{-3}$ eV.  
 The energy difference between the HOMO and  LUMO is taken
  as $\Delta E$ = 0.696 eV.
 }
\begin{center}
\begin{tabular} {ccccc}
\hline\noalign{\smallskip}
 $ $          & $\HH-\HH$     &  $\LL-\LL$   & $\HH-\LL$    &  \\
\hline\noalign{\smallskip}
 $ b1 $       & $209.3$   &  $-1.9$   & $-51.2$  &  (stacking)  \\
 $ p1(p) $    & $28.1$    &  $-12.4$  & $19.9$  & Layer 1   \\
 $ p2 $       & ---     &  ---  & $17.1 $  &      \\
\noalign{\smallskip}\hline\noalign{\smallskip}
 $ b2 $       & $49.9$   &  $-80.4$   & $-67.2$  &  (stacking)  \\
 $ q1(q) $    & $10.8$    &  $8.1$  & $9.3$  & Layer 2   \\
 $ q2 $       & ---     &  ---  & $9.2$  &      \\
\noalign{\smallskip}\hline\noalign{\smallskip}
 $ a1 $       & $-28.2$   &  $14.6$   & $-20.1$  &    \\
 $ a2 $       & $2.2$    &  $1.3$  & $-1.7$  &  Interlayer   \\
 $ c1 $       & $15.4$     &  $12.7$  & $14.1$  &      \\
 $ c2 $       & $-3.9$     &  $15.8$  & $-11.8$  &      

\\
\noalign{\smallskip}\hline
\end{tabular}
\end{center}
\label{table_1}
\end{table}

 Figure \ref{fig1} displays  the crystal structure of [Pd(dddt)$_2$]
 of a three-dimensional system 
 with eight molecular orbitals, 
which consist of four molecules (1, 2, 3, and 4)
   with HOMO and LUMO  per unit cell.
 These molecules are located on two kinds of layers, 
 where the layer 1    includes molecules 1 and 3, 
     and the layer 2 includes  molecules 2 and 4, respectively. 
The original unit cell is given by 
 lattice vectors, $\bm{a}_{\rm o}$, $\bm{b}_{\rm o}$, and $\bm{c}_{\rm o}$,
while  a new unit cell is introduced  
 by a transformation,  $ \bm{a} = -(\bm{a}_{\rm o}+\bm{c}_{\rm o})$, 
$\bm{b} = -\bm{b}_{\rm o}$, and $\bm{c} = \bm{c}_{\rm o}$
 in the present  band calculation.
Thus, the $a$ axis becomes parallel to layer 1 and layer 2.

A TB model corresponding to Fig.~\ref{fig1} 
  has been  recently obtained using the 
 crystal structure observed  under pressure.\cite{Kato2019_TB}
There are several kinds of transfer energies between two molecular orbitals,
which are listed in Table \ref{table_1}.
 The interlayer  energies in the $z$ direction  are given 
    by   $a$ (1 and 2 molecules, and 3 and 4 molecules),
             and $c$ (1 and 4 molecules, and 2 and 3 molecules). 
The intralayer  energies in the $a$-$b$ plane are given by 
   $p$ (1 and 3 molecules) and $q$ (2 and 4 molecules)
    and   $b$  (perpendicular to the $a$-$c$ plane). 
Further, these energies are classified by 
 three  kinds of  transfer energies  given by   
   HOMO-HOMO (H), LUMO-LUMO (L),  and HOMO-LUMO (HL).

 The TB model Hamiltonian   is  expressed as   
\begin{equation}
H_{\rm TB} = \sum_{i,j=1}^{N} \sum_{\alpha,\beta}
  t_{i,j;\alpha, \beta} \ket{i, \alpha} \bra{j, \beta} \; ,
\label{eq:H_model}
\end{equation}
where  
 $t_{i,j;\alpha, \beta}$ are transfer energies between nearest-neighbor sites and $\ket{i, \alpha}$ is a state vector. 
$i$ and $j$ are the  lattice sites  of the unit cell 
  with  $N$ being the total number of the unit cells, 
     $\alpha$ and $\beta$ denote the 8 molecular orbitals 
      given by HOMO $(\HH1, \HH2, \HH3, \HH4)$ and 
         LUMO $(\LL1, \LL2, \LL3, \LL4)$.  
 These  energies  in the unit of eV
   are listed in Table \ref{table_1} where 
 the gap 
 between the energy of HOMO and that of LUMO is taken 
 as $\Delta E = $ 0.696 eV to reproduce 
  the energy band of the first principle calculation.\cite{Kato2017_JPSJ}

Using a Fourier transform 
$ \ket{\alpha(\bm{k})}$ 
 $= N^{-1/2} \sum_{j} \exp[- i \bm{k}\bm{r}_j] \; \ket{j,\alpha}$
       with a wave vector  $\bk = (k_x, k_y, k_z)$,
Eq.~(\ref{eq:H_model}) is rewritten as
\begin{equation}
H_{\rm TB} = 
 \sum_{\bm{k}} \ket{\Phi(\bm{k})} \hat{H}(\bm{k}) \bra{\Phi(\bm{k})}\; , 
\label{eq:H_TB} 
\end{equation}
 where
$\bra{\Phi(\bm{k})} = (\bra{\HH1},\bra{\HH2},\bra{\HH3},\bra{\HH4}$, $\bra{\LL1}, \bra{\LL2}, \bra{\LL3}, \bra{\LL4})$. 
We take the lattice constant as unity and then 
 $0 < |k_x|, |k_y|, |k_z| < \pi$ in the first Brillouin zone. 


 The  matrix Hamiltonian $\hat{H}(\bm{k})$
  is given in Appendix A. 
The  nodal line 
   has been found using  $H(\bk)$ \cite{Kato2017_JPSJ} where the Dirac point is supported  by the existence of 
 the inversion center.\cite{Herring1937} 
The energy band  $E_j(\bk)$ 
 and the wave function $\ket{\Psi_j(\bk)}$, $(j = 1, 2, \cdots, 8)$ 
 are calculated from 
\begin{equation}
\hat{H}(\bm{k}) \ket{\Psi_j(\bk)} 
 = E_j(\bk) \ket{\Psi_j(\bk)} \; , 
\label{eq:energy_band}
\end{equation}
 where $E_1 > E_2 > \cdots > E_8$ and 
\begin{equation}
\ket{\Psi_j(\bm{k})} = \sum_{\alpha}
 d_{j,\alpha}(\bk) \ket{\alpha} \; ,
\label{eq:wave_function}
\end{equation}
 with $\alpha =$  
 H1, H2, H3, H4, L1, L2, L3, and L4. 

Since the electron close to the chemical potential is relevant 
 for the electron-hole excitation, 
 we consider only   $ E_4(\bk)$ and  $E_5(\bk)$, i.e., 
 the valence and conduction bands for the calculation 
 of the conductivity. 
Thus $E_4(\bk)$ and $E_5(\bk)$ are replaced by $E_+(\bk)$ and $E_-(\bk)$ 
 for the calculation of the conductivity, 
 while $E_{\pm}(\bk)$ represents not only the Dirac cone but also 
 full dispersion of $E_4(\bk)$ and $E_5(\bk)$ in the first Brillouin zone. 
The present energy bands $E_{\pm}(\bk)$  provide a nodal line, i.e., a loop of
 the Dirac point $\bkD$, which is obtained  from 
\begin{equation}
E_+(\bkD) = E_-(\bkD) \; .
\label{eq:Dirac_point}
\end{equation}

The chemical potential $\mu =  \mu(T)$ is determined self-consistently 
in the clean limit  from 
\begin{eqnarray}
 & & \frac{1}{N} \sum_{\bk} \sum_{j=1}^{8} f(E_{j}(\bk) - \mu(T)) 
     \nonumber \\
  = & &\int_{-\infty}^{\infty} {\rm d} \omega \; D(\omega)  f(\omega - \mu) =  4 \; ,  
 \label{eq:eq15}
\end{eqnarray}
 where 
$f(\omega)= 1/(\exp[\omega/T]+1)$ with $T$ being temperature 
 in the unit of eV 
 and $k_{\rm B }=1$.
 Equation (\ref{eq:eq15})
 is  the condition of the half-filled band 
 due to the HOMO and LUMO bands.

 $D(\omega)$ denotes a density of states (DOS) per spin and per unit cell, 
 which is given by 
\begin{eqnarray}
D(\omega) &=& \frac{1}{N} \sum_{\bk} \sum_{\gamma = \pm}
 \delta (\omega - E_{\gamma}(\bk)) \; ,
  \label{eq:dos}
\end{eqnarray}
 where  $\int {\rm d} \omega D(\omega) = 8$.

\begin{figure}
  \centering
\includegraphics[width=8cm]{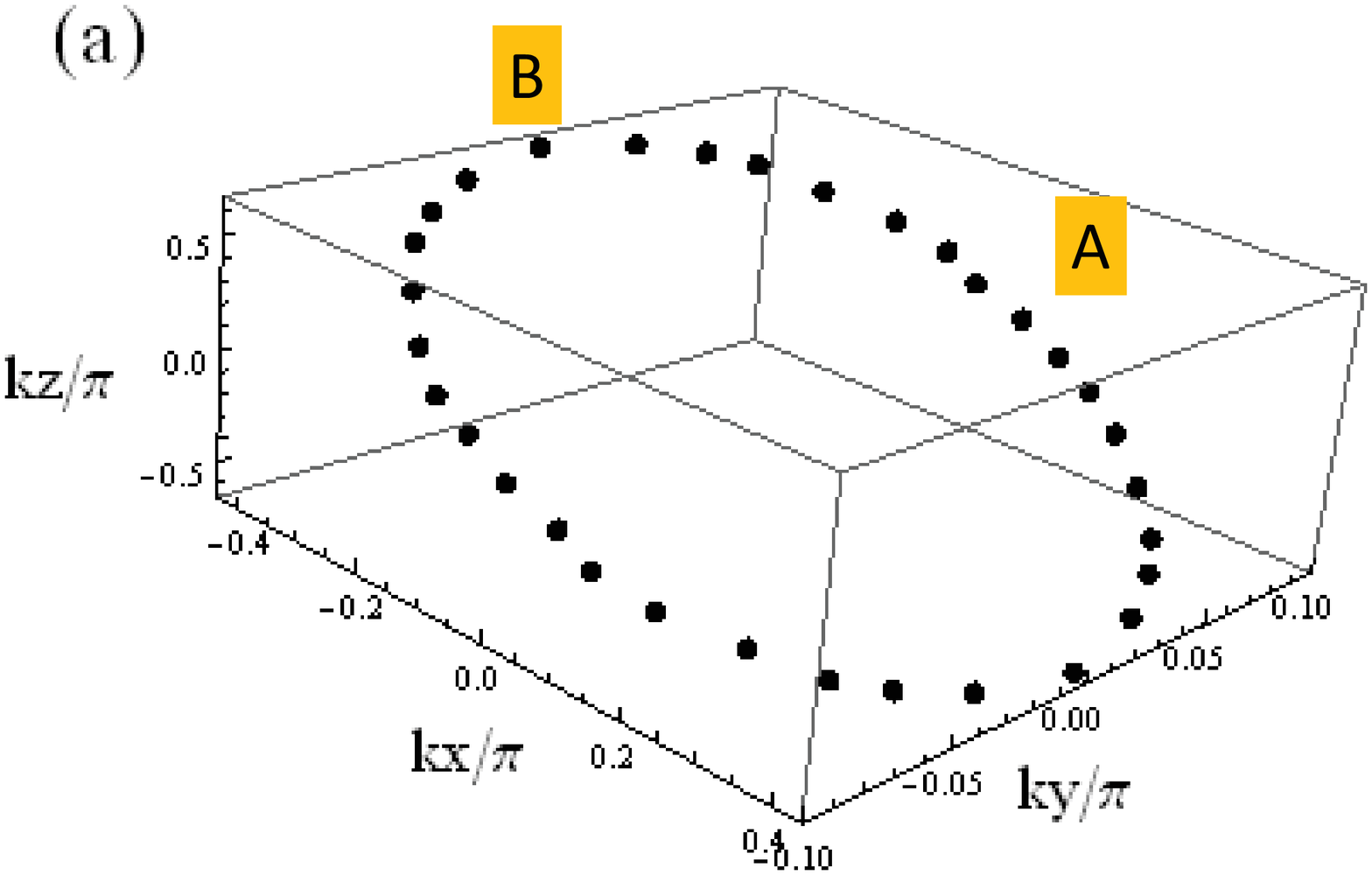} \\
\includegraphics[width=6cm]{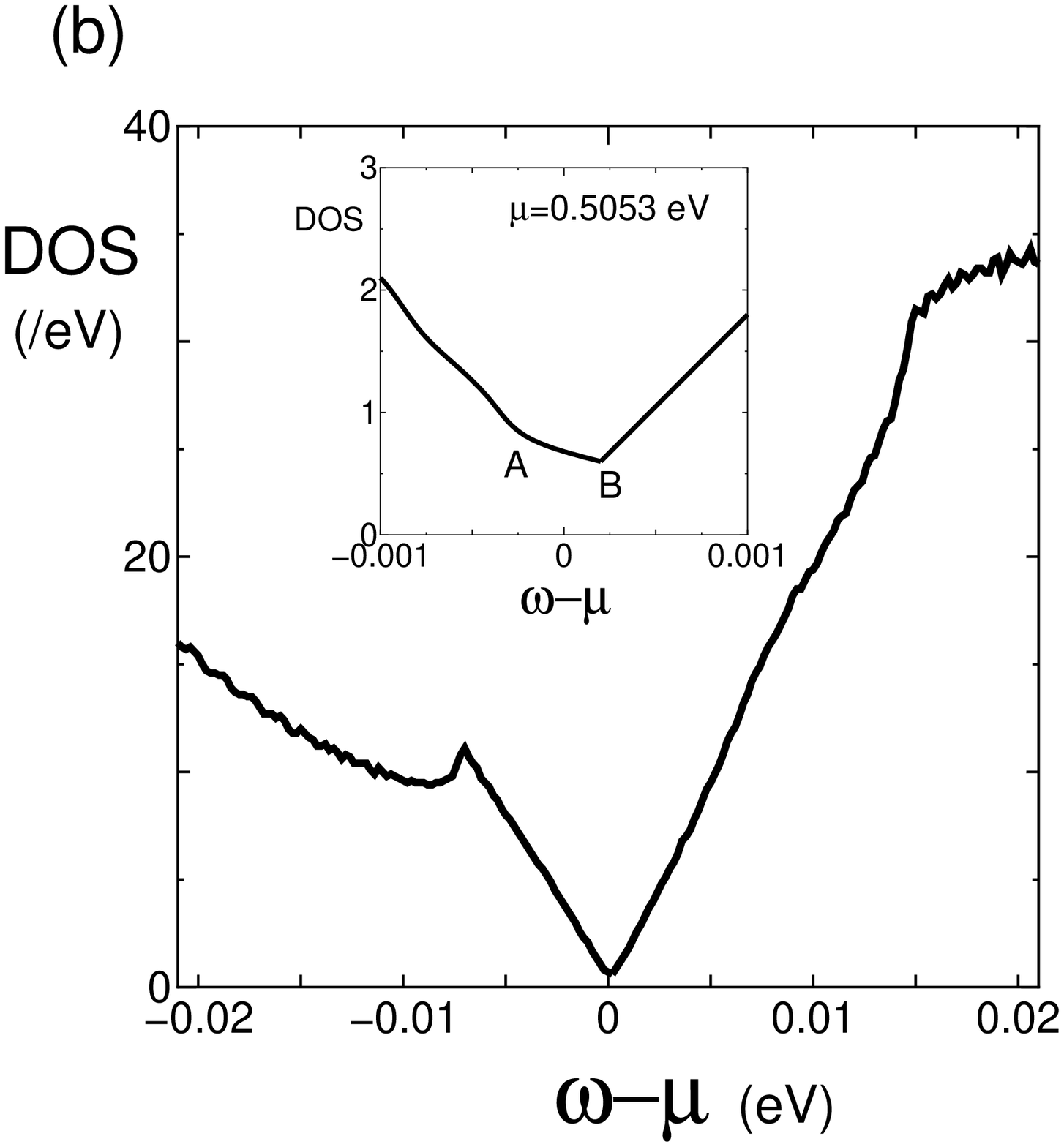}
    \caption{(Color online)
(a)Nodal line in the three-dimensional momentum space, 
 which connects  Dirac points (symbols). 
(b)Density of states (DOS), 
 as a function of $\tomega = \omega - \mu$ with 
 $\mu$ =  $\mu(0)$ = 0.5053 eV. 
The inset 
 denotes the behavior close to $\tomega = 0$, where 
 A (B) corresponds to $k_z/\pi$ = 0 ( the maximum of $|k_z/\pi|$ 
 in the loop.  
 }
\label{fig2}
\end{figure}

Figure \ref{fig2}(a)   shows a nodal line,
a loop of Dirac point $\bkD$, which is obtained 
 from Eq.~(\ref{eq:Dirac_point}). 
The Dirac point $\bkD$  is symmetric with respect to 
the $\Gamma$ point, $\bk = (0,0,0)$. 
 There  is a  mirror symmetry with respect to $k_y = 0$, which gives 
 two typical Dirac points given by (0, 0.096, 0)$\pi$  (A),  
 and (-0.42, 0, 0.64)$\pi$ (B).  
The energy $E_\pm(\bkD)$ increases from 0.5052  to 0.5056  
 as the Dirac point moves from (A) to (B) and the chemical potential 
$\mu$ = 0.5053 is found on the line between (A) and (B)   
 suggesting the nodal line semimetal close to a zero-gap state. 

Figure \ref{fig2}(b) shows the DOS, $D(\tomega)$, 
 for $-0.02 < \tomega < 0.02$,
 where the origin of $\omega$ is taken at the chemical potential
 $\mu=\mu(0)$, 
 i.e., $\tomega = \omega - \mu$.
 The peak for $\tomega > 0$ corresponds to the maximum of HOMO while 
 that of $\tomega< 0$ also exists due to the minimum of LUMO, where 
both peaks are located at $\sim \pm 0.0025$  
(not shown in the Figure). 
We note that there is a sufficient  region of $\tomega$, where 
 $D(\tomega) \propto |\tomega|$.
Such a behavior comes from two-dimensional character of the Dirac cone, 
which gives  $D(\tomega) = |\tomega|/2\pi v^2$ with $v$ being the averaged 
 velocity of the cone along the nodal line.
 From the comparison, we obtain $v \sim 0.025$, which is about a half of 
 \ET.\cite{Katayama2009_EPJ} 
 The inset denotes $D(\omega)$ close to the chemical potential, 
    which is not symmetric with respect to $\tomega$ = 0 and the minimum is located  in the region of  $\tomega > 0$. 
Compared with the previous model,\cite{Kato2017_JPSJ}
 these behaviors  are similar but   
  $D(\mu)$ of the present case is  much smaller suggesting 
    a robust Dirac cone with  
     the almost  zero-gap state.

\subsection{Total Hamiltonian}
We consider a  total Hamiltonian 
  consisting of the electron and phonon systems: 
\begin{equation}
H= H_0 + H_{p} + H_{e-p} + H_{imp} \; , 
\label{eq:H}
\end{equation}
where the first three terms denote the   Fr\"ohlich Hamiltonian 
\cite{Frohlich}  
   applied to the  Dirac electron system.
The first term $H_0$
 represents the energy band $E_{\g}(\bk)$ of
  single-component molecular conductor [Pd(dddt)$_2$],~\cite{Kato_JACS}
 and  is given by Eq.~(\ref{eq:H_model}),  
\begin{equation}
  H_0 =  \sum_{\bk} \sum_{\g = \pm} E_{\g}(\bk) a_{\g,\bk}^\dagger a_{\g,\bk}  
\; , 
\label{eq:H_0}
\end{equation}
 where 
 $E_+(\bk)$ and $E_-(\bk)$  correspond to   
  $E_4(\bk)$ and $E_5(\bk)$, respectively  in  Eq.~(\ref{eq:energy_band}).
 The spin is ignored for simplicity. 
$a_{\g,\bk}^{\dagger}$ is the creation operator of the electron with a wave vector $\bk$ and  the  $\g$ band. 
The second term $H_{p}$ describes the acoustic phonon with 
   a spectrum of $\omega_{\bq} = v_s q$ and a creation of operator 
  $b_{\bq}^{\dagger}$, where    $v \gg v_s$ with  $v$ being 
 the  velocity  of the Dirac cone. 
  The wave vector is defined by the Fourier transform 
 on the square lattice with the lattice constant taken as unity. 
ve vector $\bk$ of the  $\g$ band and acoustic phonon with a vector $\bq$. 
$\g$ = + (-) denotes a conduction (valence) band. 
The third term $H_{e-p}$  represents an electron-phonon (e-p) interaction 
 expressed as  
\begin{equation}
 H_{e-p} = \sum_{\bk, \g} \sum_{\bq}
   \alpha_{\bq} a_{\g,\bk + \bq}^\dagger a_{\g,\bk} \phi_{\bq} \; ,
\label{eq:H_int}
\end{equation}
where 
 $\phi_{\bq} = b_{\bq} + b_{-\bq}^{\dagger}$.
 Later, we introduce  a coupling constant $\lambda = |\alpha_{\bq}|^2/\omega_{\bq}$, 
  which becomes  independent of $|\bq|$  for small $|\bq|$. 
The scattering by the phonon is considered 
 within  the same band (i.e., intraband) 
  due to the energy conservation with $v \gg v_s$. 
The fourth term $H_{imp}$ denotes a normal  impurity 
 scattering, which is introduced to avoid the infinite conductivity 
in the presence of only the e-p interaction 
\cite{Holstein1964}. 
We take $k_{\rm B} = \hbar$ = 1.

\subsection{Conductivity}

The  damping of the electron of the $\g$ band, 
 $\Gamma_\g$  which 
is obtained from the Green function  expressed as,   
\begin{subequations}
\begin{eqnarray}
 G_\g(\bk, i \omega_n)^{-1} & = & 
 i \omega_n - E_{\g,\bk}+ \mu 
  + i \Gamma_{\g}  
  \; ,
 \label{eq:eq9a} \\
\Gamma_{\g} & = & \Gamma_0 + \Gamma_{\rm ph}^{\g}
 \; , 
 \label{eq:eq9b} 
  \end{eqnarray} 
\end{subequations}
 where 
$\Gamma_{\rm ph}^{\g} = - {\rm Im} \Sigma_\g (\bk, E_{\g, \bk} - \mu)$, 
  and  the real part 
 can be neglected for small doping.~\cite{Suzumura2018_PRB}
 The quantity $\Gamma_0$,  which is the damping by the impurity scattering,  
  is taken as  a parameter to scale the energy. 
Note that $\Gamma_{\rm ph}$ does not depend on $\Gamma_0$,
 and that the ratio $\Gamma_{\rm ph}^{\g}/\Gamma_0$ 
  is crucial to determine 
 the $T$ dependence of the conductivity.    
The quantity $\Sigma_\g (\bk, \omega) = \Sigma_\g (\bk, i \omega_n)$ 
 with $i\omega_n \rightarrow \omega +  0$ 
denotes 
 a self-energy of the electron Green function estimated as 
\cite{Abrikosov} 
\begin{eqnarray}
 & & \Sigma_\g (\bk, i \omega_n)  =  T \sum_m \sum_{\bq}\; |\alpha_q|^2 
            \nonumber \\
 & &\times   \frac{1}{i \omega_{n+m} - \xi_{\g, \bk+\bq}} 
      \times \frac{2 \omega_{\bq}}{\omega_{m}^2 + \omega_{\bq}^2} \; , 
 \label{eq:self_energy}
  \end{eqnarray} 
which is a product of electron and phonon Green functions. 
$\omega_n=  (2n+1)\pi T$, $\omega_{m}=2\pi m T$ with $n$ and $m$ being integers. 
$\xi_{\g, \bk} = E_{\g, \bk} - \mu$ where  $\mu$ is a chemical potential.

Here we note that 
there is a following Dirac point $\bkD$ on the nodal line. 
When  $E_{\pm}(\bk)$ contributes to a linear dispersion 
 of DOS in Fig.~\ref{fig2}(b), 
the energy dispersion $E_+(\bk)$ and $E_-(\bk)$ are given by the Dirac cone 
 as a function of $\bk - \bkD$, which 
 gives a  plane  perpendicular to a tangent of  
 the nodal line at  $\bkD$.  
\cite{Tsumuraya2018_JPSJ} 
 In this case, the damping by the acoustic phonon scattering 
 can be estimated based on the previous calculation of 
two-dimensional Dirac cone,~\cite{Suzumura2018_PRB}  
which is given by 
 $\Gamma_{\rm ph} \propto T vk$ with $ k=|\bk - \bkD|$.
Such a result is extended to   the present case 
 by  taking account of the variation of $\mu$ due to finite temperatures   
 and the  three dimensional spectrum,$ E_{\pm}(\bk)$. 
 In fact, we obtain 
\begin{subequations}
\begin{eqnarray}
  \Gamma_{\rm ph}^\g &=& CRT|\xi_{\g,\bk}|
  \; ,
 \label{eq:eq11a}
        \\ 
R &=& \frac{\lambda}{ \lambda_0} \; ,
 \label{eq:eq11b} 
\end{eqnarray}
 \end{subequations}
where  $C$ = 12.5 (eV)$^{-1}$ and $\xi_{\g, \bk} = E_{\g, \bk} - \mu$. 
The quantity $\lambda$ denotes the coupling constant of the e-p 
interaction, where 
$\lambda = |\alpha_{\bq}|^2/\omega_{\bq}$ 
 and is treated as a parameter. 
The quantity $R$ denotes  a normalization of  $\lambda$ by $\lambda_0$ , where 
  a typical value of  $\lambda_0$ =0.03 
 corresponding  to a weak coupling   is taken and  is compared with     
  the energy of the Dirac cone $\sim 0.3$.

Using $d_{\alpha \gamma}$  in Eq.~(\ref{eq:wave_function}), 
 the electric conductivity  per spin and per unit cell 
  is calculated.
The conductivity is given by\cite{Katayama2006_cond,Suzumura2017_JPSJ} 
\begin{eqnarray}
\sigma_{\nu}(T) &=&  
   \int_{- \infty}^{\infty} d \omega
   \left( - \frac{\partial f(\omega) }{\partial \omega} \right) F_{\nu}(\omega) \; ,
   \label{eq:sigma} 
       \\
  F_{\nu}(\omega) &=&  
  \frac{e^2 }{\pi \hbar N} 
  \sum_{\bk} \sum_{\gamma = \pm}  \sum_{\gamma' = \pm} 
 \overline{v^\nu_{\gamma \gamma'}(\bk)} 
  v^{\nu}_{\gamma' \gamma}(\bk) \Pi_{\g',\g} \; ,
        \nonumber \\
 \Pi_{\g',\g}&= &
     \frac{\Gamma_{\g'}}
    {(\omega - \xi_{\g',\bk})^2 +  \Gamma_{\g'}^2}
  \times
    \frac{\Gamma_{\g}}
     {(\omega - \xi_{\g,\bk})^2 +  \Gamma_{\g}^2}
  \; ,   \nonumber \\
     \label{eq:Fz}
  \\
  v^{\nu}_{\gamma \gamma'}(\bk)& = & \sum_{\alpha \beta}
 \overline{d_{\alpha \gamma}(\bk)} 
   \frac{\partial \tilde{H}_{\alpha \beta}}{\partial k_{\nu}}
 d_{\beta \gamma'}(\bk) \; ,
  \label{eq:v}
\end{eqnarray}
 where $\nu = x, y,$ and $z$, and 
 $h = 2 \pi \hbar$. $h$ and $e$ denote
   a Plank's constant and electric charge,  respectively. 
 $\xi_{\bk \gamma} = E_{\gamma}(\bk) - \mu$ 
 and $\mu$ denotes a chemical potential. 
 
 The previous calculation of the conductivity was performed 
 by taking account of only the impurity scattering of 
 the damping $\Gamma_0$,\cite{Suzumura_Kato2017_JPSJ}
 while the damping by phonon scattering
$\Gamma_{\rm ph}$\cite{Suzumura2018_PRB}
 is added to the   new TB model~
\cite{Kato2019_TB}
in the present calculation.
The total number of the lattice site  is given by $N=N_xN_yN_z$, where 
 $N_xN_y$  is the number of the intralayer sites and 
 $N_z$ is   the number of the layer.   
Note that the calculation of Eq,~(\ref{eq:sigma}) 
 with the summation of $k_z$ in the end, i.e., a two-dimensional conductivity 
 for the fixed $k_z$  
  is utilized to calculate  the nodal line semimetal as shown previously.\cite{Suzumura2017_JPSJ} 
It is noted that Eq.~(\ref{eq:sigma}) can be understood 
 using  DOS when  the intraband contribution ($\gamma=\gamma'$)
  is dominant   and $\bk$ dependence of 
 $v_{\gamma'\gamma}^{\nu}$ is small.

\section{Electric transport}
The conductor of [Pd(dddt)$_2$] under pressure exhibits 
 a Dirac electron system in  three-dimension.
The purpose of this section is to explain  
  the almost temperature independent 
  conductivity of such a system 
   by taking account of the acoustic phonon scattering.
The conductivity Eq.~(\ref{eq:sigma}) per spin is calculated, which is   
   normalized by 
 $e^2/\hbar$, i.e., $ e^2/\hbar \rightarrow 1$.
 The unit of energy is taken as eV,

We examine  the temperature $T$ dependence 
 of the conductivity  $\sigma_{\nu}$ ($\nu$ = $x$, $y$, and $z$), 
 which is determined 
  by the combined effect of impurity and   phonon scatterings. 
Both scattering decrease $\sn$  but the combined effect  is complicated 
 at finite $T$. With increasing temperature, 
 $\sn$ increases  due to the DOS of the Dirac cone but $\sn$ is suppressed  
 by the phonon scattering,  which  increases  with increasing $T$. 
We first show  the $T$ dependence of $\sn$  
 in the presence of only impurity scattering, i.e., $R$ =0. 
Next $T$ dependence of $\sn$ in the presence of both impurity and phonon 
 scatterings are shown in detail by varying  parameters $\Gamma_0$ and $R$ 
 to compare with  the $T$ dependence found in the experiment.
 The numerical calculation of the conductivity is performed 
 by choosing  $\Gamma_0$ = 0.0005 and  0.0003, which are smaller 
 than  the previous one,\cite{Suzumura_Kato2017_JPSJ}.
Hereafter, we use $\Gamma_0'= 10^{4}\Gamma_0$.

\subsection{Conductivity by impurity scattering}
\begin{figure}
  \centering
\includegraphics[width=8cm]{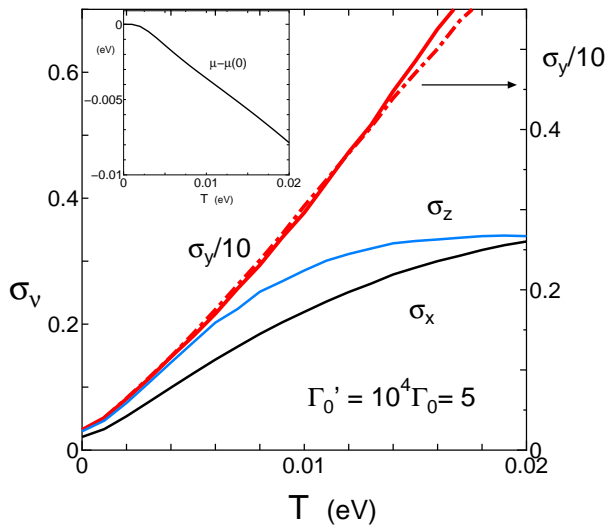}
     \caption{ (Color online)
Temperature dependence of conductivity of $\sigma_{\nu}$ 
($ \nu =x, y$, and $z$) 
with 
  $\Gamma_0'=10^4\Gamma_0 = 5$, where  
  $R$ = 0.
The dot-dashed line denotes $\sigma_{y}$,
  which is calculated 
 with the fixed $\mu =\mu(0)$= 0.5053 eV. 
The inset denotes temperature dependence of $\mu(T)$. 
  }
\label{fig3} 
\end{figure}

 In Fig.~\ref{fig3}, 
 the anisotropic conductivity   with only impurity scattering 
 is shown  by the solid line.
Compared with the result obtained by the   previous TB model,~
\cite{Suzumura_Kato2017_JPSJ} 
a noticeable result is   an  existence of 
  $T$-linear dependence of $\sigma_y$, 
     which  is seen  in a wide region of  
       temperature due to the robust Dirac cone.  
As seen from  Fig.~\ref{fig1} and Table \ref{table_1}, 
the large magnitude of $\sigma_{y}$ 
 is obtained since the molecules are stacked along the  $y$-axis. 
 Further  $\sigma_y(0)$ is  smaller than the previous one 
    due to  smaller $D(0)$.  
Such a linear increase is also found 
    for  both $\sigma_x(T)$ and  $\sigma_z(T)$, 
      in which linear increase begins to deviate    at lower temperatures. 
 There is a large  anisotropy, where   $\sigma_y$ is much larger
    than   $\sigma_x$ and $\sigma_z$.
 We note that the anisotropy of $\sigma_\nu$ comes mainly 
     from that of the velocity. 
 In fact, $v_x$ = 0.0094, $v_y$ =0.058, and $v_z$ =0.0072,
      where $v_{\nu} = \sqrt{< v_{\nu}^2 >}$ 
       and $< \cdots >$ denotes an  average by summation of $\g$, $\g'$ 
         and $\bk$ of Eq.~(\ref{eq:v}). 
Note that the velocity obtained from DOS in Fig.~\ref{fig2}(b)
     is an average of these velocities. 
However the ratio of $\sigma_{\nu}$ is different from that of velocities 
    and is more complicated, since  Eq.~(\ref{eq:sigma}) is calculated 
      as a combined effect of the velocity     and a quantity associated 
          with the electron Green function. 
The inset denotes  the $T$   dependence of 
    the chemical potential $\mu(T)$, 
    which decreases monotonously, suggesting 
       the increase of hole close to the nodal line. 
A fitting formula of $\mu$ is given by 
 $\mu - \mu(0) \simeq - 0.45 T^2/(T+ 0.003)$. 
 The dot-dashed line in the main figure 
    denotes $\sigma_y(T)$, which is calculated 
      with the fixed $\mu = \mu(0)$. 
 The difference between the solid and  dot-dashed lines 
   is negligibly small even at finite temperature. 
However such a difference  becomes noticeable  
  for a moderate  strength of the e-p interaction 
     as shown in the next subsection.  
\subsection{Effect of phonon scattering on the conductivity}
Since $\sigma_y$ is the largest one and the direction is 
 the same as that of  measured one,~\cite{Cui}
 we examine $\sigma_y(T)$ in detail with some choices of $R$. 
Before the numerical result, 
we examine $\sigma_y$  semi-analytically. 

\begin{figure}
  \centering
\includegraphics[width=8cm]{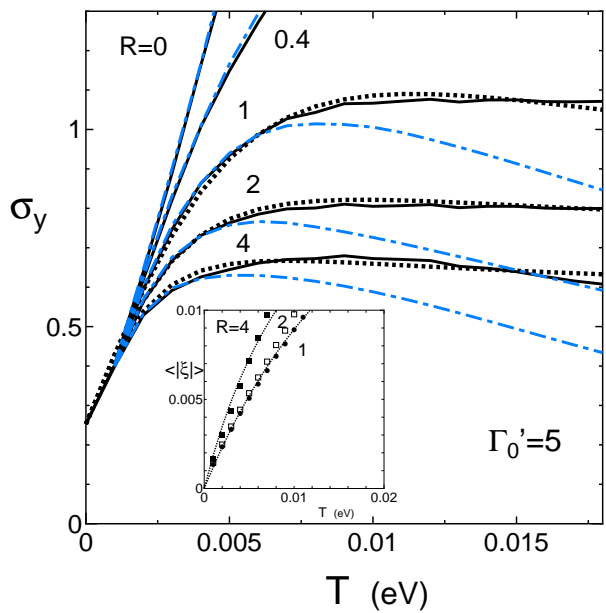} 
\caption{  (Color online)
 $T$ dependence of $\sigma_y$ with  $\Gamma_0' = 5$
    for $R$ = 0, 0.4,  1, 2, and 4 (solid line). 
 The dot line  denotes a fitting one  obtained from Eq.~(\ref{eq:eq17}).
The dot-dashed line denotes $\sigma_y$  obtained  
  for the fixed $\mu = \mu(0)$, 
   where the deviation from the solid line starts at lower $T$ 
     for the larger $R$.
The inset denotes $T$ dependence of $<|\xi|>$ obtained 
    from Eq.~(\ref{eq:eq19}) with a fitted line for $R$ = 0.5, 1, and 4.
 }
\label{fig4}
\end{figure}

\begin{figure}
  \centering
\includegraphics[width=8cm]{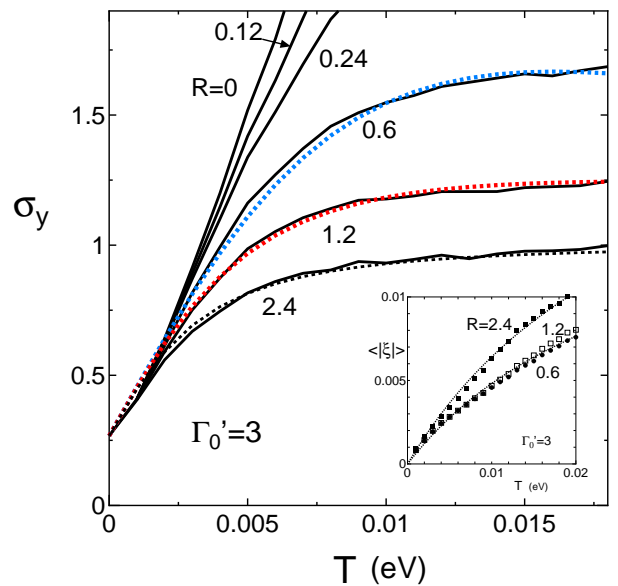} 
\caption{  (Color online)
$T$ dependence of $\sigma_y$ with  $\Gamma_0' = 3$
 for  $R$ = 0, 0.12, 0.24,  0.6, 1.2, and 2.4 (solid line). 
The dot line  denotes a fitting one  obtained from Eq.~(\ref{eq:eq17}). 
 The symbols in the inset denotes $<|\xi|>$ 
    obtained from Eq.~(\ref{eq:eq18b}),
       which are fitted by the dot line. 
}
\label{fig5}
\end{figure}

We show   $\sigma_y(T)$ can be written as, 
\begin{eqnarray}
\sigma_y(T) \simeq a + \frac{X}{1+ bX^2/(1+cX)} \; ,  
\label{eq:eq17}
\end{eqnarray}   
where $X = T/T_0$.
 Parameters $T_0, a, b$,and  $c$ in Eq.~(\ref{eq:eq17}) 
 are determined  by fitting    $\sigma_y$ 
    in Figs.~\ref{fig4} and \ref{fig5}.
Equation (\ref{eq:eq17}), which  describes well 
    for both low and high temperatures, 
      is obtained as follows.
Noting that 
  $\sigma_y$ with  only impurity scattering is written as   
     $\sigma_y = T/\Gamma$ with 
       $\Gamma \propto \Gamma_0$.~\cite{Katayama2006_cond} 
  In the presence of  $\Gamma_{\rm ph}^{\g} = \Gamma_{\rm ph}$,
 $\sigma_y$ may be written  in the same form, 
 where  $\Gamma_0$ is replaced by  
$\Gamma = \Gamma_{\g}  =  \Gamma_0 + \Gamma_{\rm ph}$,  i.e., 
\begin{subequations}
\begin{eqnarray}
   \sigma_y &=& a +  a'\frac{T}{ < \Gamma>}  \nonumber \\
    &=& a +  a'\frac{T}{ \Gamma_0 + < \Gamma_{\rm ph}>}  \nonumber \\
     & = & 
      a + \frac{T/T_0}{1+  <\Gamma_{\rm ph}>/\Gamma_0}, 
\label{eq:eq18a}
\end{eqnarray}
 where $T_0$ = $\Gamma_0/a'$. 
 Using Eq.~(\ref{eq:eq11a}) and  a relation,  
\begin{eqnarray}
<|\xi|> & = & <|\xi_{\g,\bk}|> = \frac{b'T}{1 + c(T/T_0)} \; ,
\label{eq:eq18b}
\end{eqnarray}
which is shown numerically in the inset of Fig.~\ref{fig4}, 
$<\Gamma_{\rm ph}>$  may be written as    
\begin{eqnarray}
    \frac{<\Gamma_{\rm ph}>}{\Gamma_0} 
    & = &\frac{CRT<|\xi|>}{\Gamma_0}   \nonumber \\
 &= &\frac{b (T/T_0)^2}{1+ c (T/T_0)} \; ,
\label{eq:eq18c}
\end{eqnarray}
\end{subequations}
where $b = CRb'T_0^2/\Gamma_0$.
 Thus  Eq.~(\ref{eq:eq17}) is derived  from 
 Eqs.~(\ref{eq:eq18a}),(\ref{eq:eq18b}), and (\ref{eq:eq18c}). 
As shown later,  Eq.~(\ref{eq:eq18b}) can be verified  
by calculating a  quantity 
\begin{eqnarray}
< \Gamma_{\rm ph} > = CRT < |\xi_{\g, \bm{k}}| > 
 = \frac{F}{\sigma_{y}} \; ,
\label{eq:eq19}
\end{eqnarray}   
 where $F$ is calculated from Eq.~(\ref{eq:sigma})
      with 
       $ \Pi_{\g',\g}$ replaced by  $ \Gamma_{\rm ph} \Pi_{\g',\g}$ 
         in Eq.~(\ref{eq:Fz}). 
Since $< |\xi_{\g, \bm{k}}| >$ denotes an average of energy 
   close to the nodal line , 
 we may write  $<|\xi_{\g, \bk}|> \sim T$, and 
    $< \Gamma_{\rm ph}> \sim CRT <|\xi_{\g, \bk}|>  
   \sim CRT^2 $. 
 This gives a simpler expression compared with Eq.~(\ref{eq:eq17}), i.e., 
 $\sigma_y \simeq a +  (T/T_0)/(1+ b'' (T/T_0)^2)$, 
  with $b'' \sim CRT_0^2/\Gamma_0$.

Now we examine  $\sigma_y$   
   using   Eq.~(\ref{eq:eq17}) as a fitting line.
Figure~\ref{fig4} shows $T$ dependence of $\sigma_y$ (solid line) 
    for $\Gamma_0' =\Gamma_0 \times 10^4 $ = 5
      with the fixed  $R$= 0, 0.4, 1,  2,  and 4 (solid line).
It is found that  $\sigma_y(T)$  is reduced and takes a maximum in the presence of 
 $R$, 
      while $\sigma_y(T)$ with $R$ = 0 increases monotonously. 
It is the main purpose of the present work to demonstrate  
  that  $\sigma_y(T)$ for $R$=1 and 2  
exhibit  almost $T$ independent behavior 
 in a wide temperature region of $0.01 < T < 0.02$.
 This exotic $T$ dependence  exists  for some choices of $\Gamma_0'$ and $R$,
 as shown later. 
From the dot line  corresponding to  Eq.~(\ref{eq:eq17}),
 we obtain  $T_0$ = 0.0055, $a$=0.25, and 
 $(b,c)$ = (0.5,0.22), (1.39,0.6), and (2.6,0.81) for 
   $R$ = 1, 2, and 4 respectively. 
It is verified  that $b$ increases  and  $\sigma_y$ decreases with increasing $R$. 
Note that the almost temperature indepednent $\sigma_y$ 
 comes from the existence of the $c$ term, which weakens the effect of the $b$ term. 
This corresponds to the suppression  of   $< |\xi_{\g, \bm{k}}| >/T$ 
 as shown in the inset.
In Fig.~\ref{fig4}, $\sigma_y$ with the fixed $\mu(0)$ is also shown 
 by the dot-dashed line. 
 At high temperatures,
   $\sigma_y$ is reduced noticeably due to the deviation of $\mu$ 
    from the energy of the Dirac point.
For the solid line, the $T$ dependence of 
 $<\Gamma_{\rm ph}^\g>$ in  Eq.~(\ref{eq:eq11a}) is small 
         due to  $<E_{\g}(\bk)> \sim  \mu$, which comes  
             from  the electron-hole excitation in the 
 conductivity. 
However, for the dot-dashed line, 
  $<|E_{\g}(\bk) - \mu(T)|>$ becomes 
   large due to  the fixed $\mu (0)$, 
    resulting in  the large suppression of  $\sigma_y$, 
  where the maximum is  broadened. 
With increasing $R$,
 $T_{\rm max}$ corresponding to a maximum of $\sigma_y$ 
 decreases, while  the reduction at high temperatures 
 is almost parallel.
The inset denotes $<|\xi|>$ obtained from Eq.~(\ref{eq:eq17}),
 for $R$ = 0.5, 1, and 2,  
 where the symbols are fitted by the dot lines given by  
$<|\xi|>$ 
  $ \propto T/(1+ c'(T/T_0))$
  with $c'= 0.4 (0.2)$ 
  for $R$ = 4 (1), which corresponds well to that 
    of the fitting of $\sigma_y$.

  Figure \ref{fig5} shows 
 $T$ dependence of $\sigma_y$ for  smaller $\Gamma_0'= 3$ 
   with the fixed  $R$ = 0, 0.12, 0.24, 0.6, 1.2 and 2.4 (solid line). 
The dot line  denotes a fitting one  obtained from Eq.~(\ref{eq:eq17}),
   where parameters are estimated as $T_0$ = 0.0053, $a$=0.27, and 
       $(b,c)$ = (0.15,0.06) (0.58,0.50), and (1.9,1.4) for 
         $R$ = 0.6, 1.2, and 2.4, respectively. 
 Compared  with Fig.~\ref{fig4}, we find as follows.  
A ratio of $T_0$ to $\Gamma_0$ increases with decreasing $\Gamma_0$,
 e.g. $T_0/\Gamma_0 \simeq$  10 and 16
  for $\Gamma_0'$ = 5 and 3, respectively. 
  Although $\sigma_y$ still  increases, it is expected that the almost 
constant behavior  occurs      at higher temperature 
 due to  the increase of the phonon scattering. 
 There is a quantitative  difference between $R$ = 0.12 and $R$=0.36. 
With decreasing  $\Gamma_0$,   
     $\sigma_y$ increases followed by the increase of 
   the crossover temperature from $T$ linear behavior 
    to that being almost constant. 
In the inset,  $<|\xi|>$ obtained from Eq.~(\ref{eq:eq18b}) 
    is shown by the symbols, where the dot line denotes 
 the fitting line given by 
 $<|\xi|> \propto$  
    $T/(1+ c'(T/T_0))$ with 
  $c'= 0.17 (0.18) $ for $R$ = 2.4 (0.6).

\begin{figure}
  \centering
\includegraphics[width=8cm]{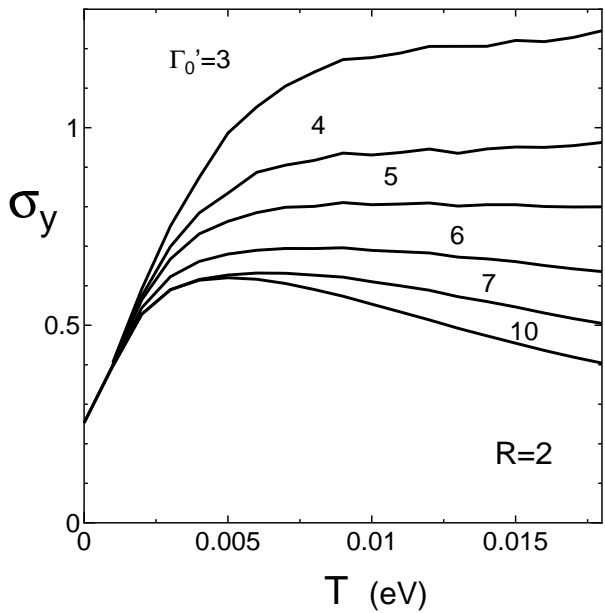}   
\caption{
$T$ dependence of $\sigma_y$, 
 for 
 $\Gamma_0'$ = 3, 4, 5, 6. 7, and 10
 with fixed $R$ =2. 
  }
\label{fig6}
\end{figure}

Figure \ref{fig6} shows $\sigma_y$  for 
some choices of  $\Gamma_0'$ with the fixed $R$ = 2, 
 where the lines with $\Gamma_0'$ = 3 and 5 
     are the same as those in Figs.~ \ref{fig5} and \ref{fig4}.
With increasing $R$, the  $T$ dependence changes  from the monotonous increase 
 to the constant behavior and a maximum is followed for $0 < T < 0.02$.
This suggests the constant behavior is obtained 
 for the moderate strength of $\Gamma_0'$ 
 in contrast to  Figs.~\ref{fig4} and \ref{fig5}, which 
 are examined for the fixed $\Gamma_0'$.
The fitting by Eq.~(\ref{eq:eq17}) is  also valid, since 
$\Gamma_0'$ = 3 and 5  are already shown in Fig.~\ref{fig5} and Fig.~\ref{fig4}. 
The line with $\Gamma_0'$ = 10 shows a noticeable decrease of $\sigma_y$ suggesting 
the dominant effect by the large e-p scattering at high temperatures. 
From Eq.~(\ref{eq:eq17}), it turns out that 
    $b$ suppresses  $\sigma_y$ but $c$ enhances  $\sigma_y$.
The role of $c$ is crucial to obtain the almost constant conductivity, 
 since a relation $bX^2/(1+cX) \propto T$ at high $T$ 
  gives almost $T$ independent $\sigma_y$
  in  Eq.~(\ref{eq:eq17}). 
We found  the almost $T$ independent $\sigma_y$.
 for 
 $(\Gamma_0',R)$  = (4, 0.8),(4,1.2), (4,2). (5,0.5), 
 (5,1), (5,2), (6,1), (6,1.2), and (6, 2).
It turns out  that the almost $T$ independent $\sigma_y$ 
 in the interval region of  $0 < T < 0.02$ 
 is realized for $4 < \Gamma_0'< 6$  and $1 < R < 2$. 

Here we discuss  the result of almost temperature independent 
conductivity in  Dirac electrons, which is obtained 
by taking account of both impurity scattering ($\Gamma_0$)
  and the acoustic phonon scattering 
($\Gamma_{\rm ph}$).
In the presence of only $\Gamma_0$, the conductivity increases 
linearly with increasing $T$ due to the increase of the relevant DOS
being  linear in  $\omega$ ( Fig. \ref{fig2}(b)).
Such  behavior is found in both the 2D Dirac cone at zero doping 
and the half-filled [Pd(dddt)$_2$]  with 
 a  nodal line, where 
 the chemical potential is located close to the nodal line 
 and the energy variation along the nodal line is very small 
 as seen from  the inset of Fig.~\ref{fig2}(b). 
However such an increase of the conductivity 
is suppressed by a moderate strength of the acoustic phonon scattering.
The  conductivity of the 2D Dirac cone shows a broad maximum as a function of $T$, 
while that of the nodal line further exhibits 
    the wide temperature region 
 followed by the almost temperature independent conductivity.
Such a result is obtained by treating the damping of phonon scattering 
 as $\Gamma_{\rm ph} \propto  T|E_{\g,\bk}-\mu |$ (Eq.~(\ref{eq:eq11a})), 
with $\mu \simeq - 0.45 T$ (the inset of Fig.~\ref{fig3})
due to asymmetry of DOS, where
 a peak around $\omega -\mu \simeq 0.03$ is much larger than 
 that of  $\omega -\mu \simeq -0.03$.
The suppression of the conductivity is understood by 
$<|E_{\g,\bk}-\mu |> \simeq T$ in the inset of Fig.~\ref{fig4}, 
 which results in  $\Gamma_{\rm ph}/\Gamma_0 \propto T^2$
due to the e-p excitation 
 in  the conductivity Eq.~(\ref{eq:sigma}).
However Eq.~(\ref{eq:eq18c}) suggests a constant conductivity by 
   the $c$-term  in the denominator.
  The wide temperature region for the relevant $c$-term could be 
 attributable to a nodal line with a half-filled band.

\subsection{Comparison with experiment}
\begin{figure}
  \centering
\includegraphics[width=7cm]{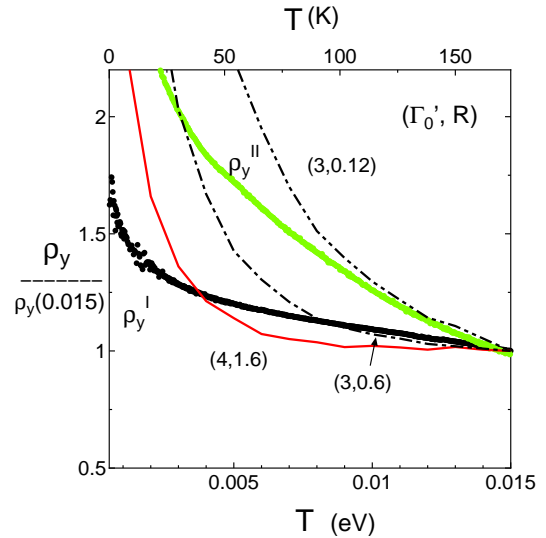}   
\caption{(Color online)
$T$ dependence of $\rho_y (= 1/\sigma_y)$
 normalized by the value at $T$=0.015 eV 
(174 K) in the interval range 
 of 0.0005 eV $< T < $0.015 eV.
The quantities $\rho^{\rm I} $ and $\rho^{\rm II}$ 
 denote the normalized resistivity  
  at $P$ = 12.6 GPa~\cite{Kato_JACS}
 and 13 GPa,~
\cite{Kato2020_tobe_submitted} respectively, 
  which are compared with 
those  of $(\Gamma_0',R) = $  
 $(4,1.6), (3, 0.6),$ and   $(3,0.12)$.
 The line (4, 1.6) stays almost constant 
 at high temperature. 
 }
\label{fig7}
\end{figure}

 Finally, in Fig.~\ref{fig7},
 we compare the present theoretical results 
 with those of the experiment.
\cite{Kato_JACS,Kato2020_tobe_submitted} 
The conductivity of a single crystal under quasi hydrostatic pressure was measured by the four-probe method using the DAC (Diamond Anvil Cell) technique.
 For the convenience of comparison, 
 we show the resistivity 
 $\rho_y (= 1 /\sigma_y)$.
Since the comparison of the absolute value is complicated
  both theoretically and experimentally, 
 we use  a simple method that the quantity is normalized at 
a  temperature of $T$ = 0.015. 
In Fig.~\ref{fig7}, the experimental result is shown by 
 $\rho_y^{\rm I}$ and $\rho_y^{\rm II}$ corresponding to 
$P$=12.6 GPa, and P=13 GPa, respectively,  which increase  about 
 1.8 and 3.4 $\times \rho$ (0.015), respectively, 
 with decreasing temperature.
The results of the present calculation are shown 
 for $(\Gamma_0', R)$ 
= (4,1.6),  (3,0.6), and (3,0.12). 
The line of (3,0.6) reproduces  
$\rho_y^{\rm I}$ well for $0.0008 < T$, 
 and  the line of (4,1.6) shows almost 
constant resistivity for  $0.0007 < T < 0.015$.
The rapid increase of $\rho_y^{\rm II}$ 
 is compared with  that of (3,0.12)
 for $0.01 < T < 0.015$, where 
 the weaker e-p coupling is taken.
Such  behavior could be expected for higher pressures due to 
the increase of the velocity.  
The larger damping  $\Gamma_0' (> 7)$ (not shown in the Figure) gives 
 a minimum for a moderate choice of $R$ with increasing $T$.
Such a  minimum, which could be relevant 
 to that of \ET~\cite{Tajima2007_EPL},  
 is obtained for a slightly large $\Gamma_0'$.

\section{Summary and Discussion}
Using a TB model, which has been recently obtained, 
we examined the temperature dependence 
of  the conductivity $\sigma_y$  
 of  [Pd(dddt)$_2$] with a nodal line semimetal.
 The main finding is  
the almost constant resistivity  obtained by the acoustic phonon scattering,  
which  plays an exotic role in the presence of the Dirac cone and nearly 
 the half-filled band.
The DOS shows linear behavior in the sufficient energy 
 region, which suggests a robust Dirac cone 
 even for the nodal line.

The conductivity $\sigma_y$ is the largest one since the transfer energy 
 is large 
  along the stacking direction of $y$-axis.
The  phonon scattering, which occurs in the half-filled Dirac electron system,
   has an  effect of reducing the linear increase of the conductivity 
 and shows  almost temperature independent  behavior at high temperatures.
With increasing temperature, the conductivity increases linearly 
 in the presence of only the impurity scattering,
 but  the presence of acoustic phonon gives the strong suppression  
 of the conductivity. The mechanism of damping  
 by  the  phonon scattering is  noticeable.   
 The damping by phonon, i.e., the imaginary  part 
 of the self-energy is proportional to $T$ $|\xi_{\g,\bk}|$ with 
 $\xi_{\g,\bk}$ is the energy of the electron 
 measured from the chemical potential.

This is in contrast to the conventional case with a Fermi surface corresponding to 
 a large doping. 
For $\mu(0) \gg T$,  $\sigma(0)$ becomes large  but $\sigma(T)$  
 as a function of $T$ decreases rapidly,~\cite{Suzumura2018_PRB}  
 due to  the large effect of  the  phonon scattering with a damping being proportional  to $T$.\cite{Sarma}

Thus it is concluded that the present  conductivity 
 being almost temperature independent
 is an evidence of the Dirac electron system with a half-filled band. 

Finally, we comment on the validity of the present TB model.
 In the previous work,\cite{Kato_JACS, Kato2017_JPSJ} 
transfer energies of the TB model have been estimated for 
 both 0 GPa and 8 GPa (corresponding to 12.6 GPa of the experiment), where
the former (latter) provides the insulating gap (Dirac points) in accordance with the experiment. Thus, the present TB model may be justified 
at lower pressures, but 
the state at higher pressure given by $P$ = 13 GPa in Fig.~\ref{fig7} shows 
a rapid increse of the normalized $\rho_y^{\rm II}$  being 
 opposite to  the conventional pressure-resistance relation
 suggesting a possible modification of the model.
Actually, it is known that higher pressure induces disorder in the molecular structure.\cite{Kato2020_tobe_submitted}

\acknowledgements
One of the authors (YS) thanks T. Tsumuraya  for useful 
 discussions on the nodal line semimetal.  
This work was supported by JSPS KAKENHI Grant Number
JP16H06346.

\appendix
\begin{table}
\caption{ Matrix elements of  Hamiltonian: 
  $ h_{\beta,\alpha} = \overline{h_{\alpha, \beta}}$.
}
\begin{center}
\begin{tabular} {ccccc}
\hline\noalign{\smallskip}
$h_{\HH,\HH}$        & $\HH1$ & $\HH2$ & $\HH3$ & $\HH4$ 
\\
\noalign{\smallskip}\hline\noalign{\smallskip}
$ \HH1$   & $h_{\HH1,\HH1}$   
        & $h_{\HH1,\HH2}$     
        & $h_{\HH1,\HH3}$      
       & $h_{\HH1,\HH4}$       
  \\
$ \HH2$ 
  & $  h_{\HH2,\HH1} $  
  & $ h_{\HH2,\HH2}$     
  & $ h_{\HH2,\HH3}$     
  & $  h_{\HH2,\HH4}$      
\\
$ \HH3$ 
 &  $h_{\HH3,\HH1}$   & $ h_{\HH3,\HH2}$ 
  &  $ h_{\HH3,\HH3}$       
   & $ h_{\HH3,\HH4}$     
\\
$ \HH4$ 
 & $ h_{\HH4,\HH1}$  & $ h_{\HH4,\HH2}$  &  $ h_{\HH4,\HH3}$  
  & $h_{\HH4,\HH4}$      
\\
\noalign{\smallskip}\hline
\hline\noalign{\smallskip}
$h_{\HH,\LL}$        & $\LL1$ & $\LL2$ & $\LL3$ & $\LL4$ 
\\
\noalign{\smallskip}\hline\noalign{\smallskip}
$ \HH1$   & $h_{H1,L1}$   
        & $h_{\HH1,\LL2}$     
        & $h_{\HH1,\LL3}$      
       & $h_{\HH1,\LL4}$       
  \\
$ \HH2$ 
  & $h_{\HH2,\LL1}$  
  & $ h_{\HH2,\LL2}$     
  & $ h_{\HH2,\LL3}$     
  & $  h_{\HH2,\LL4}$      
\\
$ \HH3$ 
 & $ h_{\HH3,\LL1}$  & $ h_{\HH3,\LL2}$ 
  &  $ h_{\HH3,\LL3}$       
   & $ h_{\HH3,\LL4}$     
\\
$ \HH4$ 
 & $ h_{\HH4,\LL1}$  & $h_{\HH4,\LL2}$  &  $ h_{\HH4,\LL3}$  
  & $h_{\HH4,\LL4}$      
\\
\noalign{\smallskip}\hline
\hline\noalign{\smallskip}
$h_{\LL,\LL}$        & $\LL1$ & $\LL2$ & $\LL3$ & $\LL4$ 
\\
\noalign{\smallskip}\hline\noalign{\smallskip}
$ \LL1$   & $h_{\LL1,\LL1}$   
        & $h_{\LL1,\LL2}$     
        & $h_{\LL1,\LL3}$      
       & $h_{\LL1,\LL4}$       
  \\
$ \LL2$ 
  &  $h_{\LL2,\LL1}$  
  & $ h_{\LL2,\LL2}$     
  & $ h_{\LL2,\LL3}$     
  & $  h_{\LL2,\LL4}$      
\\
$ \LL3$ 
 &  $h_{\LL3,\LL1}$  &  $h_{\LL3,\LL2}$ 
  &  $ h_{\LL3,\LL3}$       
   & $ h_{\LL3,\LL4}$     
\\
$ \LL4$ 
 &  $h_{\LL4,\LL1}$  &  $h_{\LL4,\LL2}$  &   $h_{\LL4,\LL3}$  
  & $h_{\LL4,\LL4}$      
\\
\noalign{\smallskip}\hline

\end{tabular}
\end{center}
\label{table_a1}
\end{table}

\section{TB model}
 As shown in Table \ref{table_a1}, the Hamiltonian is  divided into the 4 x 4 matrix, 
  $h_{{\rm H},{\rm H}}$, 
$h_{{\rm H},{\rm L}}$ ( and  
$h_{{\rm L},{\rm H}}$) , 
$h_{{\rm L},{\rm L}}$ corresponding to   H-H, H-L and L-L 
 components.   
In terms of  
$X = \e^{ik_x}$, $\bar{X}= \e^{-ik_x}$, 
$Y = \e^{ik_y}$, $\bar{Y}= \e^{-ik_y}$, and 
$Z = \e^{ik_z}$, $\bar{Z}= \e^{-ik_z}$, 
matrix elements for HOMO-HOMO (H-H) are evaluated.~\cite{Kato2019_TB} 
Here we show 
the  real matrix Hamiltonian $\tilde{H}$  
 from $\tilde{H} = U \hat{H} U^{-1}$, 
 where a 8 $\times$ 8 matrix $U$ has only the diagonal element given by 
$(U)_{\HH1,\HH1}, \cdots, (U)_{\LL4,\LL4}$
$ = (-i, -i(XYZ)^{1/2}, -i (XY)^{1/2}, -i\bZ^{1/2}$
     $1, (XYZ)^{1/2}, (XY)^{1/2}, \bZ^{1/2})$
Using 
$ \x = k_x/2$, 
$ \y = k_y/2$, and 
$ \z = k_z/2$,
Matrix elements $(\tilde{H})_{\alpha,\beta}/2 = \tH_{\alpha,\beta}$ are 
 calculated as  
\begin{eqnarray}
\tH_{\rm H1,H1}&=& \tH_{\rm H3,H3} = b1_{\rm H}\sin2\y  \nonumber 
\; , \\
\tH_{\rm H1,H2}&=& a1_{\rm H}\cos(\x-\y+\z)
                 +a2_{\rm H}\cos(\x+\y+\z)  \nonumber
\; , \\
\tH_{\rm H1,H3}&=& 2 p_{\rm H}\cos\x \cos\y \nonumber
\; , \\
\tH_{\rm H1,H4}&=& c1_{\rm H}\cos\z + c2_{\rm H}\cos(2\y+\z)
\; ,  \nonumber \\
\tH_{\rm H2,H2}&=& \tH_{\rm H4,H4} = b2_{\rm H}\cos 2\y
\; , \nonumber \\
\tH_{\rm H2,H3}&=& c1_{\rm H}\cos\z + c2_{\rm H}\cos(2\y-\z)
\; , \nonumber \\
\tH_{\rm H2,H4}&=& 2q_{\rm H}\cos\x\cos\y 
 \nonumber \; , \nonumber \\
\tH_{\rm H3,H4}&=& a1_{\rm H}\cos(\x+\y+\z) 
                 + a2_{\rm H}\cos(\x-\y+\z)
\; , \nonumber
\end{eqnarray}
for HOMO-HOMO elements,  
\begin{eqnarray}
\tH_{\rm H1,L1}&=& h_{\rm H3,L3} =  -  b1_{\rm HL} \sin \y
\; , \nonumber\\
\tH_{\rm H1,L2}&=&  - a1_{\rm HL}\sin(\x-\y+\z)
                   - a2_{\rm HL}\sin(\x+\y+z))
\; , \nonumber \\ 
\tH_{\rm H1,L3}&=& - p1_{\rm HL}\sin(\x+\y)
                   + p2_{\rm HL}\sin(-\x+\y)
\; , \nonumber\\
\tH_{\rm H1,L4}&=&  c1_{\rm HL} \sin\z 
                   +c2_{\rm HL}\sin(2\y+\z) 
\; , \nonumber\\
\tH_{\rm H2,L1}&=&  a1_{\rm HL}\sin(-\x+\y-\z)
                   +a2_{\rm HL}\sin(\x+\y+\z)
\; ,\nonumber \\ 
\tH_{\rm H2,L2}&=& h_{\rm H4,L4} = - b2_{\rm HL}\sin(2\y)
\; , \nonumber\\
\tH_{\rm H2,L3}&=& - c1_{\rm HL}\sin(\z) +c2_{\rm HL}\sin(-2\y+\z)
\; , \nonumber\\
\tH_{\rm H2,L4}&=& - q1_{\rm HL}\sin(\x+\y)-q2_{\rm HL}\sin(\x-\y)
\; , \nonumber \\
\tH_{\rm H3,L1}&=& - p1_{\rm HL}\sin(\x-\y)-p2_{\rm HL}\sin(\x+\y)
\; , \nonumber\\
\tH_{\rm H3,L2}&=&  -c1_{\rm HL}\sin(\z)+c2_{\rm HL}\sin(2\y-\z)
\; , \nonumber\\
\tH_{\rm H3,L4}&=&  a1_{\rm HL}\sin(\x+\y+\z) +a2_{\rm HL}\sin(\x-\y+\z)
\; , \nonumber \\
\tH_{\rm H4,L1}&=& - c1_{\rm HL}\sin(\z))+ c2_{\rm HL}\sin(2\y+\z)
\; , \nonumber\\
\tH_{\rm H4,L2}&=&  q1_{\rm HL}\sin(\x-\y) + q2_{\rm HL}\sin(\x+\y) 
\; ,\nonumber \\ 
\tH_{\rm H4,L3}&=& - a1_{\rm HL}\sin(\x+\y+\z) + a2_{\rm HL}\sin(\x-\y+\z))
\; , \nonumber 
\end{eqnarray}
for HOMO-LUMO elements,  
\begin{eqnarray}
\tH_{\rm L1,L1}&=& h_{\rm L3,L3}= \Delta E/2 + b1_{\rm L}\cos(2\y)
\; , \nonumber\\
\tH_{\rm L1,L2}&=&  a1_{\rm L}\cos(\x-\y+\z)+a2_{\rm L}\cos(\x+\y+\z)
\; , \nonumber\\
\tH_{\rm L1,L3}&=&  2 p_{\rm L}\cos\x\cos\y
\; , \nonumber\\
\tH_{\rm L1,L4}&=&  c1_{\rm L}\cos(\z)  + c2_{\rm L}\cos(2\y+\z)
\; , \nonumber\\
\tH_{\rm L2,L2}&=& h_{\rm L4,L4} =  \Delta E/2 + b2_{\rm L}\cos(2\y)
\; ,\nonumber \\
\tH_{\rm L2,L3}&=& -c1_{\rm L}\cos(\z) - c2_{\rm L}\cos(2\y-\z) 
\; , \nonumber\\
\tH_{\rm L2,L4}&=&  2 q_{\rm L}\cos\x\cos\y 
\; , \nonumber\\
\tH_{\rm L3,L4}&=& -a1_{\rm L}\cos(\x+\y+\z) -a2_{\rm L}\cos(\x-\y+z) 
\; , \nonumber  
\end{eqnarray}
for LUMO-LUMO elements,  
and $h_{\alpha,\beta} = h_{\beta,\alpha}$.
Note that matrix elements of HOMO-HOMO and LUMO-LUMO are the even function of
$\bk$ while those of of HOMO-LUMO is the  odd function due to the 
 difference in the symmetry of HOMO and LUMO around the Pd atom. 
These parities are used for constructing   2 $\times$ 2 effective Hamiltonian, 
where the diagonal element show the 
  even function   
and the off-diagonal element shows the  odd function 
 with respect to $\bk \rightarrow -\bk$.
 These relations  provide ingredients for obtaining  
  a nodal line and  Dirac points.


\end{document}